# Superconductivity at 38 K at an electrochemical interface between an ionic liquid and FeSe$_{0.8}$Te$_{0.2}$ on various substrates

Shunsuke Kouno,[1] Yohei Sato,[1] Yumiko Katayama,[1] Ataru Ichinose,[2] Daisuke Asami,[1] Fuyuki Nabeshima,[1] Yoshinori Imai,[3] Atsutaka Maeda,[1] and Kazunori Ueno[1]*

[1]Department of Basic Science, University of Tokyo, Meguro, Tokyo 115-8902, Japan

[2]Central Research Institute of Electric Power Industry, Yokosuka, Kanagawa 240-0196, Japan

[3]Department of Physics, Tohoku University, Sendai 980-8578, Japan

*Correspondence to: ueno@phys.c.u-tokyo.ac.jp

**ABSTRACT**

Superconducting FeSe$_{0.8}$Te$_{0.2}$ thin films on SrTiO$_3$, LaAlO$_3$ and CaF$_2$ substrates were electrochemically etched in an ionic liquid, DEME-TFSI, electrolyte with a gate bias of 5 V. Superconductivity at 38 K was observed on all substrates after the etching of films with a thickness greater than 30 nm, despite the different $T_c$ values of 8 K, 12 K and 19 K observed before etching on SrTiO$_3$, LaAlO$_3$ and CaF$_2$ substrates, respectively. $T_c$ returned to its original value with the removal of the gate bias. The observation of $T_c$ enhancement for these thick films indicates that the $T_c$ enhancement is unrelated to any interfacial effects between the film and the substrate. The sheet resistance and Hall coefficient of the surface conducting layer were estimated from the gate bias dependence of the transport properties. The sheet resistances of the surface conducting layers of the films on LaAlO$_3$ and CaF$_2$ showed identical temperature dependence, and the Hall coefficient was found to be almost independent of temperature and to take values of -0.05 to -0.2 m$^2$/C, corresponding to 4-17 electrons per FeSe$_{0.8}$Te$_{0.2}$ unit cell area in two dimensions. These common transport properties on various substrates suggest that the superconductivity at 38 K appears in the surface conducting layer as a result of an electrochemical reaction between the surface of the FeSe$_{0.8}$Te$_{0.2}$ thin film and the ionic liquid electrolyte.

FeSe is an iron-based superconductor with the simplest possible composition and exhibits superconductivity at 8.5 K.[1] FeSe has recently attracted considerable attention owing to the enhancements in the superconducting transition temperature ($T_c$) that can be achieved through various methods. The $T_c$ of a one-unit-cell FeSe thin film on a SrTiO$_3$ substrate has been reported to take values of 105 K and 85 K based on an in situ resistivity measurement and a diamagnetic measurement, respectively.[2,3] Spectroscopic studies of monolayer and several-layer FeSe on SrTiO$_3$ have revealed superconducting gaps corresponding to 65 K and 80 K by means of angle-resolved photoemission spectroscopy (ARPES) and scanning tunneling microscopy (STM), respectively.[4–6] These values are much higher than the bulk $T_c$ values of all known iron-based superconductors. A $T_c$ of 48 K has also been observed in several-layer FeSe on SrTiO$_3$ with carrier doping by K ions.[7] This $T_c$ enhancement has been suggested to originate from charge transfer from the oxide substrate to the ultrathin FeSe film.[6] $T_c$ enhancements of up to approximately 40 K have also been reported following the insertion of cations or a (Li$_{0.8}$Fe$_{0.2}$)OH layer to the FeSe mother compound.[8–14] Recently, electrostatic carrier doping on ultrathin FeSe films and flakes has also been found to enhance $T_c$ up to approximately 40 K.[15–19] The authors of these studies employed an electric double layer transistor (EDLT) configuration with an ionic liquid, diethylmethyl(2-methoxyethyl)ammonium bis(trifluoromethylsulfonyl)imide (DEME-TFSI), as a gate electrolyte for tuning the high-density carriers.[20–22] Since no additional phase other than FeSe was found in the X-ray diffraction (XRD) data,[18] the $T_c$ enhancement was concluded to originate from electrostatic carrier doping. In addition, a thickness dependence study conducted by means of the electrochemical etching of FeSe showed that $T_c$ enhancement occurred only when the film was thinner than 15 nm,[16,19] probably indicating that the interface between the film and the substrate also plays an important role in the $T_c$ enhancement.

We have previously reported $T_c$ enhancements in FeSe$_{1-x}$Te$_x$ thin films on various substrates.[23–26] The observed $T_c$ values depend on both the Te content and the substrate material. For example, the $T_c$ values for FeSe$_{0.8}$Te$_{0.2}$ films on CaF$_2$ and LaAlO$_3$ substrates are enhanced to 20 K and 12 K, respectively, in contrast to the $T_c$ of 8 K observed for FeSe$_{0.8}$Te$_{0.2}$ on SrTiO$_3$. This can be explained by differences in the a-axis lattice constants of FeSe$_{0.8}$Te$_{0.2}$ films on different substrates.[24] In this paper, we report the enhancement of $T_c$ up to 38 K for thick FeSe$_{0.8}$Te$_{0.2}$ films on various substrates prepared via EDLT fabrication with the ionic liquid DEME-TFSI. By means of electrochemical etching with the ionic liquid,[16] the film thickness was varied. The application of a gate bias resulted in the formation of a surface conducting layer with a $T_c$ of 38 K; with the removal of the gate bias, the surface conducting layer disappeared, causing $T_c$ to return to its original value. We also estimated the transport properties of the surface conducting layer. The surface conducting layer exhibited electron conduction with a common dependence of the mobility on temperature on various substrates.

## Results

### Characterization of $FeSe_{0.8}Te_{0.2}$ thin films

The film thickness and crystal quality were examined via XRD measurements. Figure 1 (a) shows the XRD patterns of $FeSe_{0.8}Te_{0.2}$ thin films fabricated on $LaAlO_3$ (LAO), $CaF_2$ and $SrTiO_3$ (STO) substrates. All samples exhibited (001), (002) and (004) peaks, while the (003) peak for the film on the LAO substrate was obscured by a (002) peak of the substrate. The c-axis lattice constants for the films on the LAO, $CaF_2$ and STO substrates were 5.69, 5.72 and 5.70 A, respectively, consistent with previous reports.[23,24] The full widths at half maximum (FWHMs) of the rocking curves for the (001) peak were 0.4, 0.7 and 1 deg. for the films on the LAO, $CaF_2$ and STO substrates, respectively; these values are also almost the same as those reported previously (0.2-0.6 deg.) [23], demonstrating that all of these films consisted of high-quality single-phase samples. X-ray reflectivity (XRR) measurements revealed clear thickness fringes for all films, as shown in Fig. 1 (b), indicating a smooth surface and a sharp interface between the film and the substrate. In addition, all XRR curves were well fitted by the model structure, and we estimated the thicknesses as shown in Fig. 1 (b). The film thicknesses were also confirmed by atomic force microscopy (AFM) images of the films.

### Superconducting properties of pristine and etched $FeSe_{0.8}Te_{0.2}$ thin films

The superconducting properties of the samples when subjected to gating and etching were examined for a gate bias ($V_G$) of 5 V. As shown in Fig. 2 (a), the EDLT samples were patterned in Hall bars with six electrodes, and a Pt film was placed alongside each Hall-bar-instrumented sample to act as a gate electrode. Figures 2 (c) and (d) show the temperature ($T$) dependences of the sheet resistance ($R_S$) for samples of $FeSe_{0.8}Te_{0.2}$ films on LAO and $CaF_2$ substrates (LAO and $CaF_2$ samples), respectively. First, the $R_S$-$T$ curves of the pristine sample before etching were measured for $V_G$ values of of 0 V and 5 V. Then, the temperature was increased to 250 K with a $V_G$ of 5 V while monitoring the gate current ($I_G$), as shown in Fig. 2 (b). The channel was electrochemically etched, and the drain current ($I_D$) was gradually decreased. The product of $I_G$ and time is a Faradaic charge ($Q_F$) that is proportional to the amount of charge of the reacted ions. After etching, $T$ was decreased while maintaining $V_G$ = 5 V, and the $R_S$–$T$ curve was measured. After several cycles of etching, the $V_G$ dependence of the $R_S$-$T$ curve for the etched sample was measured via the following procedure: First, the $R_S$–$T$ curve was measured for a $V_G$ of 5 V. Then, the temperature was increased to 250 K without $V_G$, and the $R_S$–$T$ curve for $V_G$ = 0 V was measured. Finally, a $V_G$ of 5 V was again applied at 250 K, and the $R_S$–$T$ curve for $V_G$ = 5 V was measured again. We show the $R$-$T$ curves of the pristine and etched samples for $V_G$ = 5 V, $V_G$ = 0 V, and $V_G$ = 5 V. The $R_S$–$T$ curves of the pristine samples remained almost unchanged between the $V_G$ values of 0 V and 5 V. In contrast, $T_c$ was enhanced after several cycles of etching at $V_G$ = 5 V. $T_c$ increased from 12 K to 38 K for the LAO sample and from 19 K to 38 K for the CaF$_2$ sample. As shown in S. Fig 3 in the supplementary information, critical magnetic field at 0 K was also enhanced from 47 T to 67 T on films on the LAO substrate. The coherence length at 0 K was 3.8 nm to 3.1 nm. With the removal of $V_G$, $R_S$ increased, and $T_c$ returned to the value of the pristine sample. With the application of a $V_G$ of 5 V, $R_S$ slightly decreased, and $T_c$ was enhanced to 38 K for both samples. Notably, the $R_S$ value at $V_G$ = 5 V after the removal of the initial $V_G$ was larger than that before the removal of $V_G$. Since the sheet resistance due to electrostatic carrier doping should always have the same value at $V_G$ = 5 V, this difference suggests that the change in $R_S$ can be attributed to some other origin than electrostatic carrier doping. In addition, the $T_c$ for $V_G$ = 0 V was different for the LAO and CaF$_2$ sample because of different lattice constant of the FeSe$_{0.8}$Te$_{0.2}$ films. If the surface conducting layer is produced by an electrostatic carrier doping, it is natural that the surface conducting layer also show different $T_c$ values for $V_G$ = 5 V. Then, the $T_c$ enhancement to 38 K for both samples also suggests that the origin of the carrier doping is not the electrostatic doping. This will be discussed later. Figures 2 (e) and (f) show the $T$ dependences of the Hall coefficient ($R_H$) for the same samples shown in Figs. 2 (c) and (d), respectively. $R_H$ was almost zero above 80 K and showed an increase with decreasing temperature below 40 K for the pristine samples. In contrast, $R_H$ was always negative at all temperatures for the etched samples at $V_G$ = 5 V. The $R_H$-$T$ behavior returned to almost the original one after the application of $V_G$ = 0 V. These results indicate that the etching at $V_G$ = 5 V resulted in the formation of a conducting layer on the surface, with a $T_c$ of 38 K, for both the LAO and CaF$_2$ samples. Since $R_H$ was negative, electron conduction dominated in the conductive layer. In addition, since the transport properties of the pristine and etched samples were almost the same for a $V_G$ of 0 V, it can be concluded that the surface conductive layer disappeared with the removal of $V_G$ from the etched sample.

**Thickness dependence of the superconducting properties of FeSe$_{0.8}$Te$_{0.2}$ thin films**

The thickness dependence of the superconducting properties was also examined for the LAO, CaF$_2$ and STO samples. Figures 3 (a), (b), and (c) show the $R_S$-$T$ curves of the LAO, CaF$_2$ and STO samples, respectively, with various thicknesses (numbers of etching cycles). $R_S$ is normalized to the $R_S$ value at 100 K. For the LAO and CaF$_2$ samples, $T_c$ was enhanced after several cycles of etching. The CaF$_2$ sample showed a two-step superconducting transition during the initial stage of etching. A similar two-step transition has been reported for an EDLT configuration on an FeSe flake with a gate bias of approximately 4 V and has been ascribed to the inhomogeneity of the carrier distribution.[15] In addition, the normalized $R_S$-$T$ curves after the $T_c$ enhancement were nearly identical. In contrast, for the STO sample, $T_c$ was gradually enhanced from 8 K to 38 K over many cycles. We estimated the onset temperature ($T_c^{onset}$) of superconductivity from the $R_S$-$T$ curve as shown in Fig. 3(a). We also

estimated the thickness after *n* cycles of etching, *thickness(n)*, from the following equation:

$$thickness(n) = thickness(XRR) \times \sum_{i=1}^{n} Q_F / \sum_{i=1}^{n\_total} Q_F,$$

where $Q_F$ is the Faradaic charge for each cycle, *thickness(XRR)* is the film thickness before etching as estimated from the XRR measurement, and $n_{total}$ is the total number of cycles needed for the etching of the entire film. After the total etching of the film, the sample resistance is larger than MOhm. In addition, the film totally disappeared after the etching experiment. Therefore, we assumed the entire film was etched during the $n_{total}$ cycles of etching. Figure 3 (d) shows the thickness dependence of $T_c^{onset}$ for the LAO, CaF$_2$ and STO samples. $T_c$ started to increase only after two cycles of etching and saturated at 38 K after four cycles for the LAO and CaF$_2$ samples. As reported in the previous paragraph, $T_c$ decreased to its original value with the removal of $V_G$ at 250 K, but it returned to 38 K after the next application of $V_G = 5$ V for the next cycle. The STO sample showed $T_c$ enhancement from 8 K to 16 K at a thickness of 30 nm and a further $T_c$ enhancement to 38 K at a thickness of 10 nm. Since $T_c$ enhancement was observed for thick samples on all substrates, and since $T_c$ changed with the application and removal of $V_G$, we conclude that this $T_c$ enhancement was not affected by the interface between the substrate and the film but instead originated from the surface conducting layer produced by $V_G = 5$ V.

The thickness of the samples which showed the $T_c$ enhancement was confirmed by means of transmission electron microscopy (TEM) and XRD measurements. We performed corresponding etching experiments using other samples on LAO, CaF$_2$ and STO substrates and terminated the etching process after several cycles. All of the etched samples showed superconductivity at $T_c$ values above 34 K. The film thickness after etching was directly obtained via TEM measurements. The thickness dependences of $T_c^{onset}$ for these samples are shown in Fig. 3 (d). The FeSe$_{0.8}$Te$_{0.2}$ films on LAO, CaF$_2$ and STO substrates all showed $T_c$ enhancement for films with thicknesses greater than 30 nm. As shown in Figs. 4 (a) and (b) and in S. Figs. 1 (a) and (e) in the supplementary information, the interface between the substrate and the film was smooth for all films, and a clear periodicity of the atomic arrays was observed. The bright region at the interface probably indicates Se diffusion from the film into the substrate.[27,28] For the film on the LAO substrate, the clear periodicity remained at the surface, and no additional layer was found on the film, as shown in Fig. 4 (a). On the CaF$_2$ and STO substrates, a disordered FeO$_x$ layer was found on the ordered region with clear periodicity. The $T_c$ enhancement probably occurred in this ordered region. XRD patterns recorded before and after etching indicated that the peak position remained unchanged and that no new peak was present after etching, as shown in Fig. 4 (c); the only observed difference was a decrease in the peak intensity. In addition, the FWHM of the rocking curve for the (001) peak remained unchanged, as shown in Fig. 4 (d). These XRD data indicated that no new phase was created in the film by the etching process. Thus, no electrochemical reaction occurred in the bulk of the film; instead, such reactions took place only at the surface.

We also examined the thickness dependence of $T_c$ for FeSe films on LAO and STO substrates. As shown in S. Figs. 2 (a) and (b), these films showed $T_c$ enhancements of up to 30 or 40 K upon etching. On STO substrates, only films with thicknesses below 12 nm showed $T_c$ enhancement. This finding coincides with those of previous reports.[16] In contrast, a film with a thickness of 30 nm on the LAO substrate showed $T_c$ enhancement, similar to the behavior of $FeSe_{0.8}Te_{0.2}$ films. Notably, several $FeSe_{0.8}Te_{0.2}$ samples on $CaF_2$ and STO substrates showed $T_c$ enhancement to above 37 K only for thicknesses below 10 nm. The different critical thicknesses for FeSe and $FeSe_{0.8}Te_{0.2}$ films on different substrates were probably due to differences in the homogeneity of the films. As shown in S. Figs. 1 (a) and (e) in the supplementary information, a disordered region was observed in the $FeSe_{0.8}Te_{0.2}$ film on STO. In addition, a disordered Fe (or $FeO_x$) layer was observed on top of the $FeSe_{0.8}Te_{0.2}$ films on $CaF_2$ and STO substrates. These TEM images indicate that the film quality depends on the substrate and that the best quality is achieved for $FeSe_{0.8}Te_{0.2}$ films on LAO substrates. We consider that good film quality throughout the entire film is necessary for the occurrence of $T_c$ enhancement for a thick film. Thus, although we did not perform TEM measurements of all of these samples, the lack of $T_c$ enhancement to 38 K for the thick films on some samples might have been due to insufficient film homogeneity, especially near the surface.

**Discussion**

Finally, we discuss the origin of the $T_c$ enhancement. The $T_c$ enhancement has been reported to be due to charge accumulation on the surface of the FeSe.[15,16,18] However, an electrochemically reacted layer on the surface may also show a high $T_c$, since FeSe samples intercalated with alkali ions and/or organic molecules present $T_c$ values above 40 K.[8,9] To distinguish electrostatic charge accumulation from electrochemical reaction, we estimated the sheet resistance and Hall coefficient of the surface layer. The resistance tensor, $\rho$, and the conductance tensor, $\sigma$, are represented by the following equations:

$$\rho = \begin{pmatrix} R_S & -R_H B \\ R_H B & R_S \end{pmatrix} \qquad (1)$$

$$\sigma = \rho^{-1} = \frac{1}{R_S^2 + R_H^2 B^2} \begin{pmatrix} R_S & R_H B \\ -R_H B & R_S \end{pmatrix} \sim \begin{pmatrix} \frac{1}{R_S} & \frac{R_H B}{R_S} \\ -\frac{R_H B}{R_S} & \frac{1}{R_S} \end{pmatrix} \qquad (2)$$

where $B$ is the magnetic field applied during the Hall measurement. The $\sigma$ of a sample at $V_G$ = 5 V is equal to the sum of the $\sigma$ values of the sample at $V_G$ = 0 V and of the surface conducting layer produced by a $V_G$ of 5 V. Therefore, the sheet resistance and Hall coefficient of the surface conducting layer, $R_S^{surface}$ and $R_H^{surface}$, obey the following equations:

$$\frac{1}{R_{xx}(V_G = 5V)} = \frac{1}{R_{xx}(V_G = 0V)} + \frac{1}{R_{xx}^{surface}} \qquad (3)$$

$$\frac{R_H(V_G = 5V)}{R_{xx}(V_G = 5V)} = \frac{R_H(V_G = 0V)}{R_{xx}(V_G = 0V)} + \frac{R_H^{surface}}{R_{xx}^{surface}} \qquad (4)$$

As shown in Figs. 5 (b), (c) and (d), we examined the changes in the sheet resistance and Hall coefficient for one $V_G$ cycle of a LAO sample (LAO-1, as shown in Fig. 2 (c)) and two $V_G$ cycles of a CaF$_2$ sample (CaF$_2$-1 and CaF$_2$-2, where the data for CaF$_2$-1 are shown in Fig. 2 (d)). Figure 5 (a) shows the temperature dependence of $R_S^{surface}$ normalized to the value at 90 K. The $R_S$-$T$ curves for the LAO and CaF$_2$ samples just before the last etching cycle, $R_S$(last), are also plotted. All curves follow almost the same profile. This indicates that the transport properties, such as the electron mobility and scattering time, of all samples exhibited identical temperature dependences. As shown in the inset of Fig. 5 (a), $R_H$ was almost independent of temperature and negative for all samples. Electron-type conduction is a common feature in previous reports on the $T_c$ enhancement of FeSe,[7,15–18] and the vanishing of the hole pocket at the Fermi level has been considered to be the origin of the $T_c$ enhancement.[7,15,16] The observed Hall coefficient values of 0.05 to 0.2 m$^2$/C correspond to 4-17 electrons per unit cell in two dimensions (0.376 nm×0.376 nm). If such a high density of carriers were electrostatically accumulated on the surface, then electrons would be strongly scattered at the surface, and the scattering time should change with the variation in the accumulated carrier density.[29] However, no such change was observed in the $R$-$T$ curves. In addition, as shown previously in Figs. 2 (c) and (d), the removal of $V_G$ irreversibly changed $R_S$, suggesting that the origin of the change in $R_S$ is some other phenomenon than electrostatic carrier doping. Therefore, a different cause of carrier doping other than the electric field effect is likely responsible for the $T_c$ enhancement.

Both the irreversible change in $R_S$ with $V_G$ and the lack of variation in the electron mobility with the carrier doping can be explained by assuming that the surface conducting layer is formed not by the accumulation of electrostatic charge but by an electrochemical reaction between the FeSe$_{0.8}$Te$_{0.2}$ and the ionic liquid. We hypothesize that the etching of the film and the formation of the surface conducting layer occurred simultaneously with the application of the $V_G$ of 5 V. One potential candidate of the forming reaction of the surface conducting layer is an electrochemical intercalation of DEME$^+$ ion,

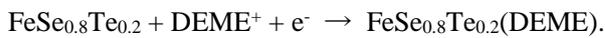
FeSe$_{0.8}$Te$_{0.2}$ + DEME$^+$ + e$^-$ → FeSe$_{0.8}$Te$_{0.2}$(DEME).

Since observed $Q_F$ during the reaction was much larger than $Q_F$ needed for this reaction, other electrochemical reaction, such as electrochemical decomposition of DEME-TFSI and dissolution of

FeSe$_{0.8}$Te$_{0.2}$, also occurs. We discussed on the possible electrochemical reactions in the supplementary information. In addition, we hypothesize that when this $V_G$ was removed, the surface conducting layer disappeared, probably due to decomposition or peeling off from the surface. Then, the abrupt decrease in the sheet conductance occurred with the removal of $V_G$. In addition, both the electron mobility and the volume charge carrier density should be identical among LAO-1, CaF$_2$-1 and CaF$_2$-2. This hypothesis was also supported by the change in the sheet conductance with the repeated etching of the LAO and CaF$_2$ samples, as shown in Fig. 5 (d). The sheet conductance at 50 K decreased with the removal of $V_G$ and, with repeated etching, gradually increased after this reduction. This behavior can be explained by an increase in the thickness of the surface conducting layer with repeated etching. If the conductance of the surface conducting layer is higher than that of the bulk FeSe$_{0.8}$Te$_{0.2}$ film at 50 K, then repeated etching will increase the sheet conductance at low temperatures. As the number of etching cycles increases, the ratio of the thickness of the surface conducting layer to the total film thickness will increase. Then, just before the film is totally removed, the surface conducting layer will cover almost the entire film. Consistent with this picture, the temperature dependences of the sheet resistance just before the last etching cycle for both the LAO and CaF$_2$ samples were also identical to that for the surface conducting layer, as shown in Fig. 5 (a). We also examined two dimensionality of the superconductivity on the surface conducting layer. When the superconducting layer is thinner than the superconducting coherence length, it behaves as a two dimensional superconductor. However, as shown in S. Fig 4 in the supplementary information, the surface conducting layer did not behave as a two dimensional superconductor. This suggests that superconducting layer is electrochemically formed and relatively thick.

**Conclusion**

In conclusion, a surface conducting layer with a $T_c$ of 38 K was formed with the electrochemical etching of FeSe$_{0.8}$Te$_{0.2}$ thin films on LAO, CaF$_2$ and STO substrates. Since the thicknesses of all etched samples with $T_c$ values of 38 K were greater than 30 nm, the enhancement of $T_c$ cannot be related to any interaction between the film and the substrate. In addition, $T_c$ enhancement was also observed for an FeSe thin film on a LAO substrate with a thickness of approximately 30 nm. The surface conducting layer again showed almost identical temperature dependences between the sheet resistance and the Hall coefficient. This finding suggests that the surface conducting layer is formed not by the accumulation of electrostatic charge on the FeSe$_{0.8}$Te$_{0.2}$ surface but by an electrochemical reaction between the FeSe$_{0.8}$Te$_{0.2}$ and the ionic liquid electrolyte. Hall coefficient measurements showed that the surface conducting layer contained 4-17 electrons per unit cell in two dimensions, with an overall negative charge. From TEM measurements, we could observe a smooth interface between the substrate and the film and a clear periodicity of the atomic arrays in the etched FeSe$_{0.8}$Te$_{0.2}$ film on the LAO substrate. These observations indicate that the formation of the surface conducting layer did not affect the bulk region of the film, and the surface conducting layer completely disappeared with the removal of $V_G$. We consider that previous studies on carrier doping in ultrathin FeSe film can be classified into two groups: those that show a $T_c$ of approximately 40 K for a several-layer FeSe film and those that show a $T_c$ above 65 K for a monolayer FeSe on STO. Our results indicate that the electrochemical doping of FeSe and FeSe$_{0.8}$Te$_{0.2}$ can result in the formation of a superconducting layer with a $T_c$ of approximately 40 K and that no interfacial interaction is necessary for the enhancement of $T_c$ to 40 K. However, we think that the interface between FeSe and STO is probably essential for the enhancement of $T_c$ above 65 K. We believe that it will be possible to prepare monolayer FeSe with a $T_c$ above 65 K with further study of the electrochemical etching of FeSe.

**Methods**

FeSe and FeSe$_{0.8}$Te$_{0.2}$ thin films were deposited by means of pulsed laser deposition (PLD) with a KrF eximer laser and an FeSe or FeSe$_{0.8}$Te$_{0.2}$ polycrystalline target. The fabrication conditions are described in detail elsewhere.[30,31] We used a commercially available STO (001) substrate with a step-and-terrace surface, a LAO (001) substrate and a CaF$_2$ (001) substrate. AFM and XRD measurements were carried out prior to device fabrication. Au(100 nm)/Ti (20 nm) films were formed via electron-beam evaporation at a base pressure of 10$^{-5}$ Torr. Since FeSe$_{0.8}$Te$_{0.2}$ thin films can be damaged by exposure to high temperatures (above 100 deg. Celsius) and water during the standard processes of photolithography and dry etching, we employed a sandblasting technique at room temperature for the fabrication of Hall bar electrodes. The films were coated with a dry film resist patterned via photolithography and were etched by sandblasting with alumina emery (#220). The Hall bar electrodes and wires were coated with a silicone sealant to prevent electrochemical reactions between the electrolyte and the Au/Ti electrode. The ionic liquid DEME-TFSI was dropped onto the channel area of the Hall bar configuration and the Pt film. Electrochemical etching and transport measurements were carried out in a He atmosphere with a Quantum Design Physical Property Measurement System (PPMS) at temperatures from 300 K to 2 K.

**References**

1. Hsu, F.-C. *et al.* Superconductivity in the PbO-type structure α-FeSe. *Proc. Natl. Acad. Sci.* **105,**

    14262–14264 (2008).

2. Ge, J.-F. *et al.* Superconductivity above 100 K in single-layer FeSe films on doped SrTiO$_3$. *Nat.*

*Mater.* **14,** 285–289 (2015).

3. Sun, Y. *et al.* High temperature superconducting FeSe films on SrTiO$_3$ substrates. *Sci. Rep.* **4,** 6040 (2014).

4. He, S. *et al.* Phase diagram and electronic indication of high-temperature superconductivity at 65 K in single-layer FeSe films. *Nat. Mater.* **12,** 605–610 (2013).

5. Qing-Yan, W. *et al.* Interface-Induced High-Temperature Superconductivity in Single Unit-Cell FeSe Films on SrTiO$_3$. *Chin. Phys. Lett.* **29,** 037402 (2012).

6. Tan, S. *et al.* Interface-induced superconductivity and strain-dependent spin density waves in FeSe/SrTiO$_3$ thin films. *Nat. Mater.* **12,** 634–640 (2013).

7. Miyata, Y., Nakayama, K., Sugawara, K., Sato, T. & Takahashi, T. High-temperature superconductivity in potassium-coated multilayer FeSe thin films. *Nat. Mater.* **14,** 775–779 (2015).

8. Burrard-Lucas, M. *et al.* Enhancement of the superconducting transition temperature of FeSe by intercalation of a molecular spacer layer. *Nat. Mater.* **12,** 15–19 (2013).

9. Hatakeda, T. *et al.* New Alkali-Metal- and 2-Phenethylamine-Intercalated Superconductors $A_x(C_8H_{11}N)_yFe_{1-z}Se$ (A = Li, Na) with the Largest Interlayer Spacings and $T_c$ ~ 40 K. *J. Phys. Soc. Jpn.* **85,** 103702 (2016).

10. Wang, A. F. *et al.* Superconductivity at 32 K in single-crystalline $Rb_xFe_{2-y}Se_2$. *Phys. Rev. B* **83,**

060512 (2011).

11. Zhang, Y. *et al.* Nodeless superconducting gap in $A_x$Fe$_2$Se$_2$ (A=K,Cs) revealed by angle-resolved photoemission spectroscopy. *Nat. Mater.* **10,** 273–277 (2011).

12. Ying, T. P. *et al.* Observation of superconductivity at 30~46K in $A_x$Fe$_2$Se$_2$ (A = Li, Na, Ba, Sr, Ca, Yb, and Eu). *Sci. Rep.* **2,** 426 (2012).

13. Krzton-Maziopa, A. *et al.* Synthesis of a new alkali metal–organic solvent intercalated iron selenide superconductor with T c ≈ 45 K. *J. Phys. Condens. Matter* **24,** 382202 (2012).

14. Lu, X. F. *et al.* Coexistence of superconductivity and antiferromagnetism in (Li$_{0.8}$Fe$_{0.2}$)OHFeSe. *Nat. Mater.* **14,** 325–329 (2015).

15. Lei, B. *et al.* Evolution of High-Temperature Superconductivity from a Low-$T_c$ Phase Tuned by Carrier Concentration in FeSe Thin Flakes. *Phys. Rev. Lett.* **116,** 077002 (2016).

16. Shiogai, J., Ito, Y., Mitsuhashi, T., Nojima, T. & Tsukazaki, A. Electric-field-induced superconductivity in electrochemically etched ultrathin FeSe films on SrTiO$_3$ and MgO. *Nat. Phys.* **12,** 42–46 (2016).

17. Hanzawa, K., Sato, H., Hiramatsu, H., Kamiya, T. & Hosono, H. Key Factors for Insulator -Superconductor Transition in FeSe Thin Films by Electric Field. *IEEE Trans. Appl. Supercond.* **27,** 1–5 (2017).

18. Hanzawa, K., Sato, H., Hiramatsu, H., Kamiya, T. & Hosono, H. Electric field-induced

superconducting transition of insulating FeSe thin film at 35 K. *Proc. Natl. Acad. Sci.* **113,** 3986–3990 (2016).

19. Shiogai, J., Miyakawa, T., Ito, Y., Nojima, T. & Tsukazaki, A. Unified trend of superconducting transition temperature versus Hall coefficient for ultrathin FeSe films prepared on different oxide substrates. *Phys. Rev. B* **95,** 115101 (2017).

20. Ueno, K. *et al.* Electric-field-induced superconductivity in an insulator. *Nat Mater* **7,** 855–858 (2008).

21. Yuan, H. *et al.* High-Density Carrier Accumulation in ZnO Field-Effect Transistors Gated by Electric Double Layers of Ionic Liquids. *Adv. Funct. Mater.* **19,** 1046–1053 (2009).

22. Ueno, K. *et al.* Field-Induced Superconductivity in Electric Double Layer Transistors. *J. Phys. Soc. Jpn.* **83,** 032001 (2014).

23. Imai, Y., Sawada, Y., Nabeshima, F. & Maeda, A. Suppression of phase separation and giant enhancement of superconducting transition temperature in FeSe$_{1-x}$Te$_x$ thin films. *Proc. Natl. Acad. Sci. U. S. A.* **112,** 1937–1940 (2015).

24. Imai, Y. *et al.* Control of structural transition in FeSe$_{1-x}$Te$_x$ thin films by changing substrate materials. *Sci. Rep.* **7,** 46653 (2017).

25. Tsukada, I. *et al.* Epitaxial Growth of FeSe$_{0.5}$Te$_{0.5}$ Thin Films on CaF2 Substrates with High Critical Current Density. *Appl. Phys. Express* **4,** 053101 (2011).

26. Nabeshima, F., Imai, Y., Hanawa, M., Tsukada, I. & Maeda, A. Enhancement of the superconducting transition temperature in FeSe epitaxial thin films by anisotropic compression. *Appl. Phys. Lett.* **103,** 172602 (2013).

27. Ichinose, A. *et al.* Microscopic analysis of the chemical reaction between Fe(Te, Se) thin films and underlying $CaF_2$. *Supercond. Sci. Technol.* **26,** 075002 (2013).

28. Tsukada, I., Ichinose, A., Nabeshima, F., Imai, Y. & Maeda, A. Origin of lattice compression of $FeSe_{1-x}Te_x$ thin films on $CaF_2$ substrates. *AIP Adv.* **6,** 095314 (2016).

29. Sato, Y., Doi, K., Katayama, Y. & Ueno, K. Electrolyte dependence of transport properties of $SrTiO_3$ electric double layer transistors. *Jpn. J. Appl. Phys.* **56,** 051101 (2017).

30. Imai, Y. *et al.* Superconductivity of $FeSe_{0.5}Te_{0.5}$ Thin Films Grown by Pulsed Laser Deposition. *Jpn. J. Appl. Phys.* **49,** 023101 (2010).

31. Imai, Y. *et al.* Systematic Comparison of Eight Substrates in the Growth of $FeSe_{0.5}Te_{0.5}$ Superconducting Thin Films. *Appl. Phys. Express* **3,** 043102 (2010).

**Acknowledgment**

This work was supported in part by JSPS KAKENHI (Grant Numbers 25708039 and 25220604) and CREST-JST.

**Author Contributions**

S.K., Y.K. and K.U. designed the research and analyzed the data. S.K. and Y.S. contributed to the device fabrication and the measurements of the transport properties and device characteristics. A.I. contributed to the TEM measurements. D.A., F.N., Y.I., and A.M. contributed to the film fabrication and to the XRD and AFM measurements. The text and figures of the paper were prepared by S.K., K.U. and A.M. All authors contributed to discussing the results reported in the manuscript.

## Additional Information

**Supplementary information** accompanies this paper.

**Competing Interests:** The authors declare no competing interests.

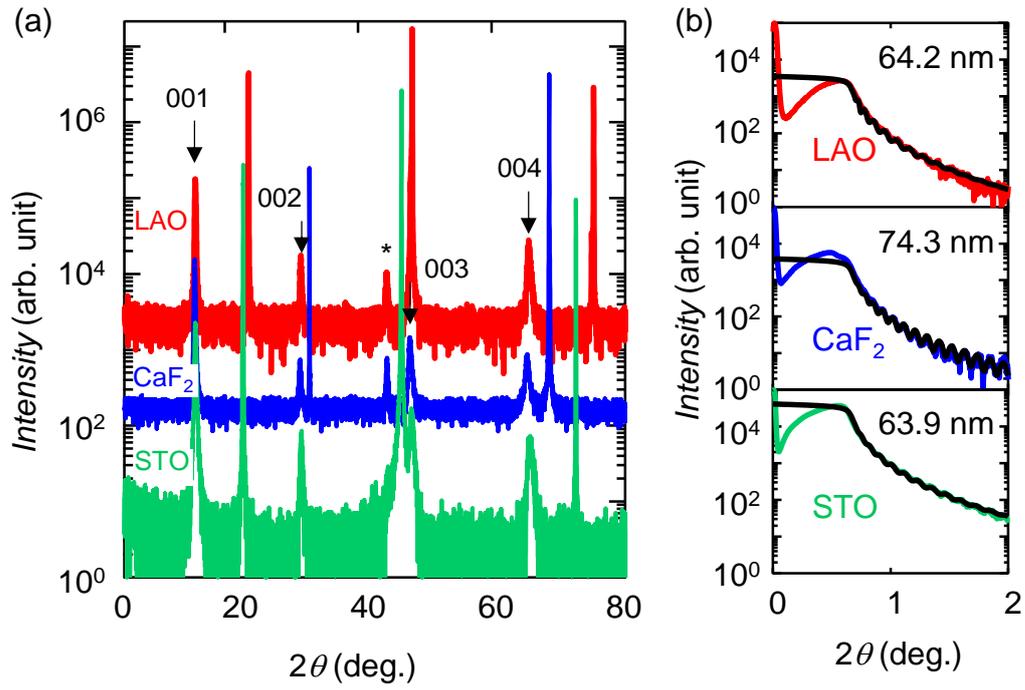

Figure 1 (a) X-ray diffraction (XRD) patterns of $FeSe_{0.8}Te_{0.2}$ thin films on $LaAlO_3$ (LAO), $CaF_2$ and $SrTiO_3$ (STO) substrates. (b) X-ray reflectivity (XRR) patterns of the films. The black lines represent fit curves. The estimated thickness is shown in each panel.

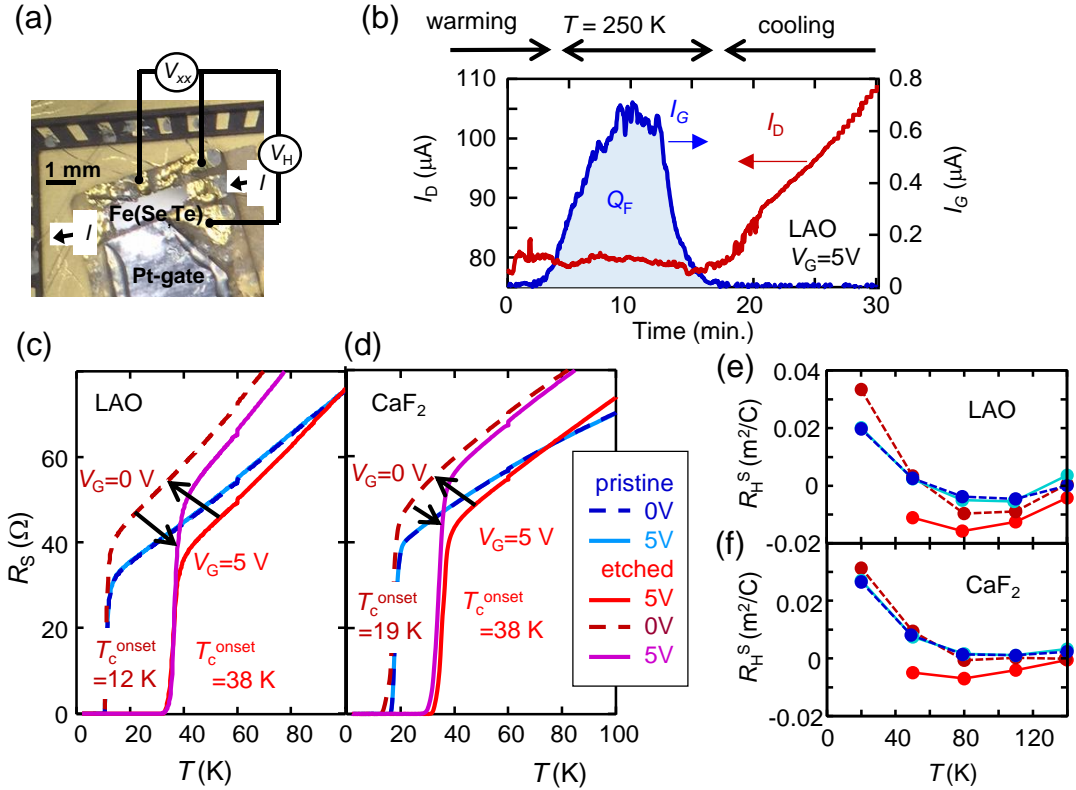

Figure 2 Changes in the superconducting and transport properties of FeSe$_{0.8}$Te$_{0.2}$ samples subjected to electrochemical etching and gating. (a) A photographic image of the electric double layer transistor (EDLT) device on the FeSe$_{0.8}$Te$_{0.2}$/LAO sample. The four-terminal resistance and the Hall resistance were simultaneously measured with the Hall bar electrodes. The ionic liquid electrolyte was placed between the film and the Pt gate. (b) The time dependences of the drain current ($I_D$) and the gate current ($I_G$) during electrochemical etching at $V_G = 5$ V and 250 K for the FeSe$_{0.8}$Te$_{0.2}$/LAO sample. $I_G$ corresponds to the Faradaic current, and the product of $I_G$ and time corresponds to the Faradaic charge ($Q_F$). (c), (d) The temperature ($T$) dependences of the sheet resistance ($R_S$) for the FeSe$_{0.8}$Te$_{0.2}$ samples fabricated on the LAO and CaF$_2$ substrates, respectively. Each panel shows the $R_S$-$T$ curves with and without gating for the pristine and etched samples. For each etched sample, bias voltages $V_G$ of 5 V (red solid line), 0 V (red broken line), and 5 V (purple solid line) were applied in that order. For each pristine sample, bias voltages $V_G$ of 0 V (blue broken line) and 5 V (blue solid line) were applied in that order. (e), (f) The $T$ dependences of the Hall coefficient ($R_H$) for the samples on the LAO and CaF$_2$ substrates, respectively. Each panel shows the $R_H$-$T$ curves with and without gating for the pristine and etched samples.

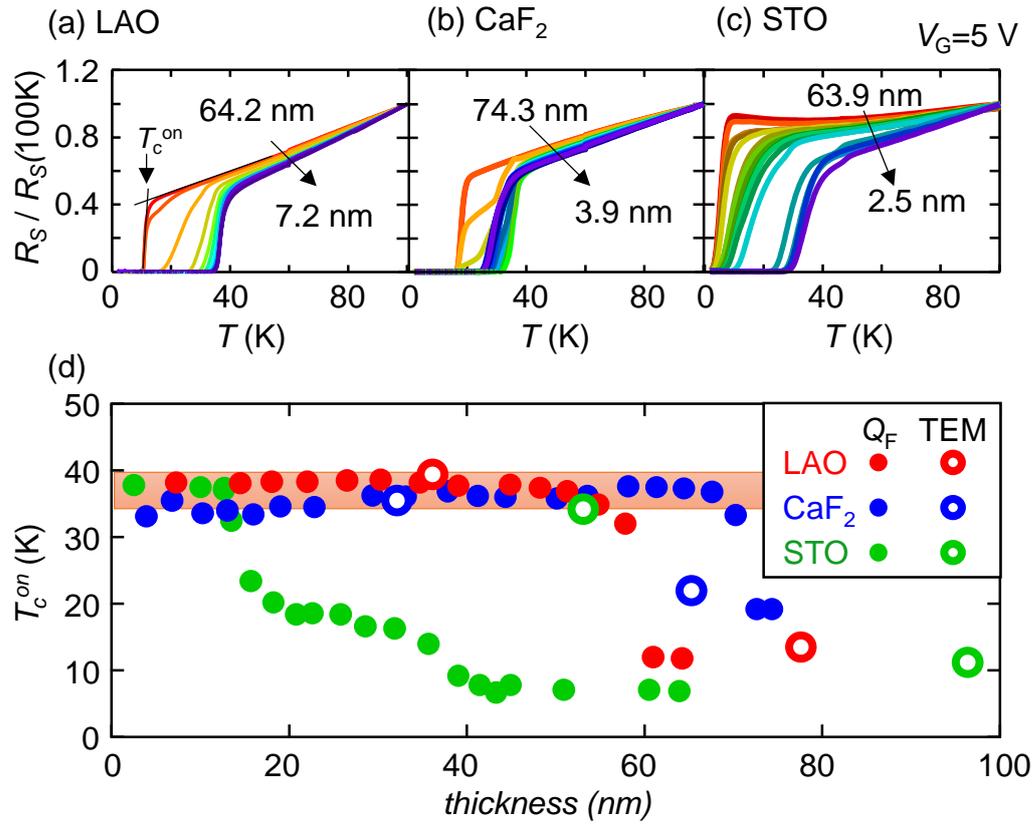

Figure 3 Variations in the superconducting transition temperature with changes in thickness for FeSe$_{0.8}$Te$_{0.2}$ samples on various substrates. (a)-(c) The $T$ dependences of $R_S$ (normalized to the $R_S$ value at 100 K, $R_S$(100 K)) for LAO, CaF$_2$ and STO samples, respectively, subjected to electrochemical etching at $V_G$ = 5 V. (d) The thickness dependences of $T_c^{onset}$ for the three samples. Filled symbols correspond to the data shown in (a)-(c), for which the thickness was estimated from $Q_F$. Open symbols correspond to other samples whose thicknesses after etching were directly measured via TEM, as shown in Fig. 4.

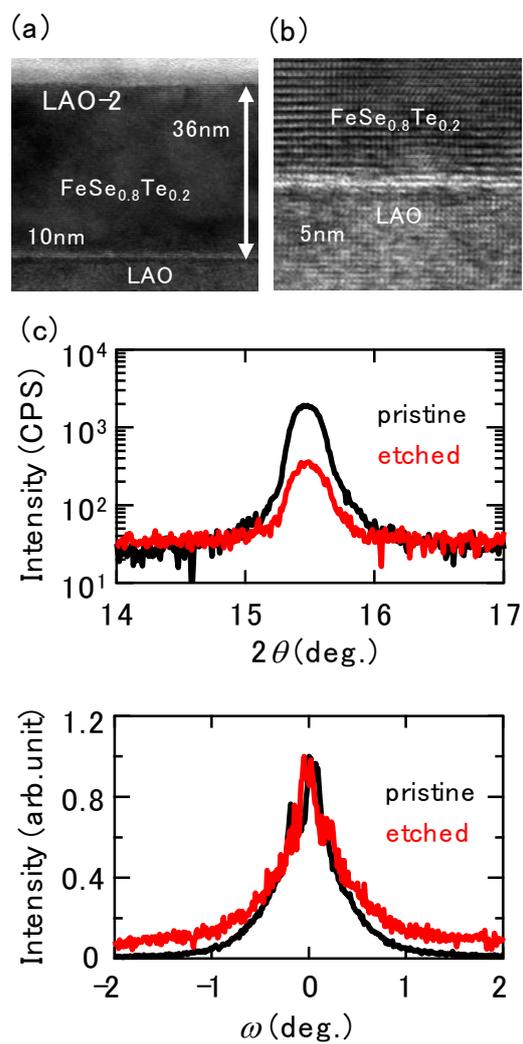

Figure 4 TEM and XRD data for the same sample on a LAO substrate before (pristine) and after (etched) electrochemical etching. (a), (b) TEM images of the etched sample. (c), (d) XRD patterns and rocking curves, respectively, for the (001) diffraction peak of the $FeSe_{0.8}Te_{0.2}$ film. The intensity is normalized to the intensity at $\omega = 0$.

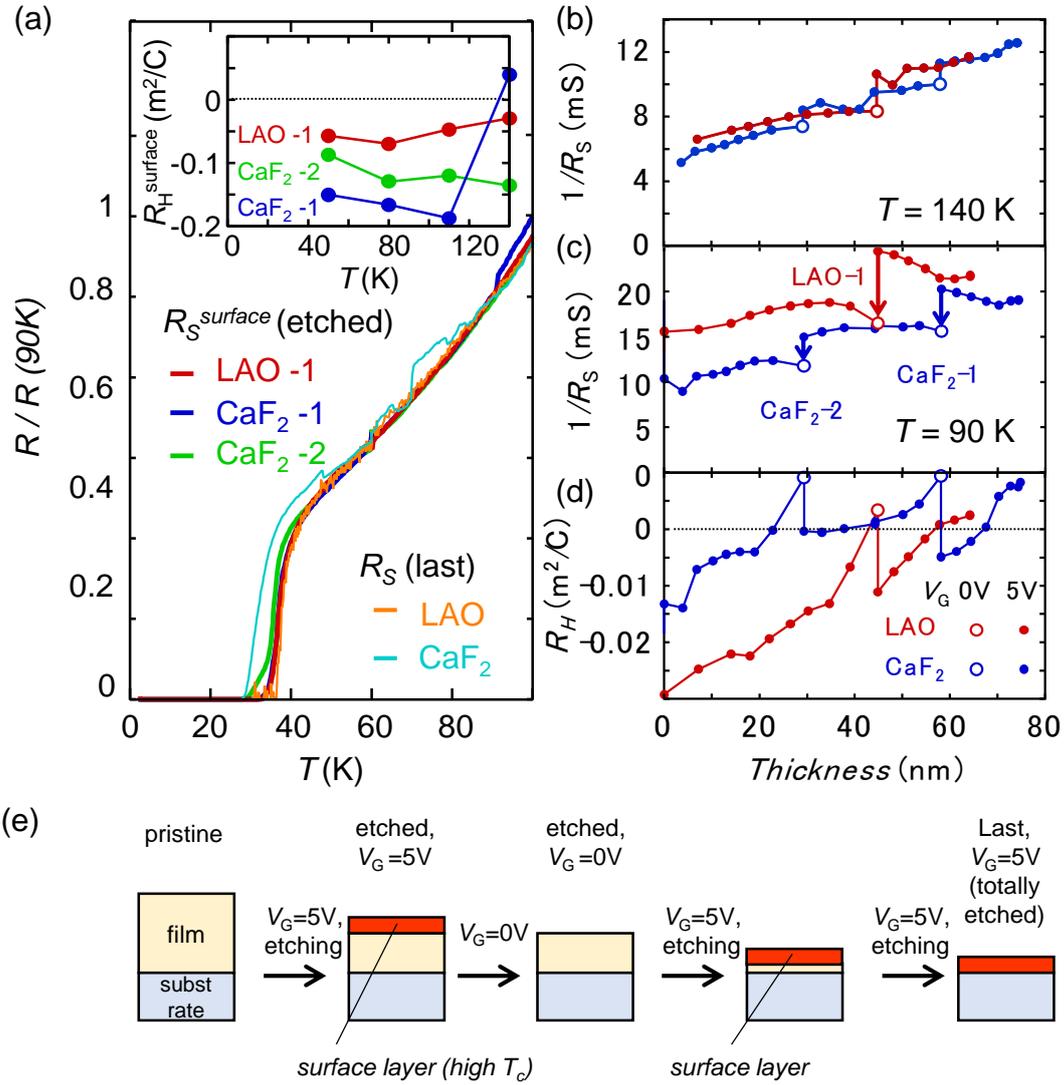

Figure 5 Superconducting transition for the surface conducting layer and changes in the transport properties with electrochemical etching for various samples. (a) The $T$ dependence of the $R_S$ values of the surface conducting layer ($R_S^{surface}$), normalized to the value at 90 K. $R_S^{surface}$ was estimated from the change in the sheet conductance between $V_G$ values of 5 V and 0 V. Data for three $V_G$ cycles of two samples, labeled as LAO-1, CaF$_2$-1 and CaF$_2$-2 in Fig. 5(c), are shown. The $R_S$-T curves just before the last etching cycle for the LAO and CaF$_2$ samples are also plotted. The inset shows the $R_H$ values of the surface conducting layers of the LAO-1, CaF$_2$-1 and CaF$_2$-2 samples as a function of temperature. (b), (c), (d) Values of $1/R_S$ at 140 K and 50 K and of $R_H$ at 50 K, respectively, as functions of the film thickness for the LAO and CaF$_2$ samples. $V_G$ was changed from 5 V to 0 V once and twice during the etching of the LAO and CaF$_2$ samples, respectively. (e) Schematic illustration of the evolution of a sample during etching. A surface layer is formed during etching and disappears with the removal of the gate bias.

# Supplementary Information

# Superconductivity at 38 K in an electrochemical interface between ionic liquid and FeSe$_{0.8}$Te$_{0.2}$ on various substrates

Shunsuke Kouno,[1] Yohei Sato,[1] Yumiko Katayama,[1] Ataru Ichinose,[2] Daisuke Asami,[1] Fuyuki Nabeshima,[1] Yoshinori Imai,[3] Atsutaka Maeda,[1] and Kazunori Ueno[1]*

[1]Department of Basic Science, University of Tokyo, Meguro, Tokyo 115-8902, Japan

[2]Central Research Institute of Electric Power Industry, Yokosuka, Kanagawa 240-0196, Japan

[3]Department of Physics, Tohoku University, Sendai 980-8578, Japan

## TEM images after etching for samples on CaF$_2$ and SrTiO$_3$

Scanning TEM experiments were performed on FeSe$_{0.8}$Te$_{0.2}$ films on CaF$_2$ and STO substrates. $T_c$ was changed from 22.0 K to 35.5 K for the film on the CaF$_2$ substrate and 11.2 K to 34.K for the film on the STO substrate after several cycles of the etching. Topographic images for both films include a bright area with a thickness of several atomic layers at the interface between the substrate and the film, probably indicating Se diffusion from the film into the substrate. In addition, both films include a bright, disordered area on the top of a dark area with a clear periodic atomic arrays. Since O and Fe atoms are observed but Se atoms are not observed in the disordered area, this area corresponds to the oxidized Fe (FeO$_x$) layer. In addition, there are no FeO$_x$ peaks from the $2\theta$-$\theta$ measurement by the XRD before and after the etching, the FeO$_x$ is a polycrystal or an amorphous. The FeO$_x$ layer possibly originates from chemical reaction between air and a residual Fe layer after the electrochemical etching of FeSe. We estimated a film thickness from the dark, well-crystallized FeSe area and plotted in Fig. 3 in the manuscript.

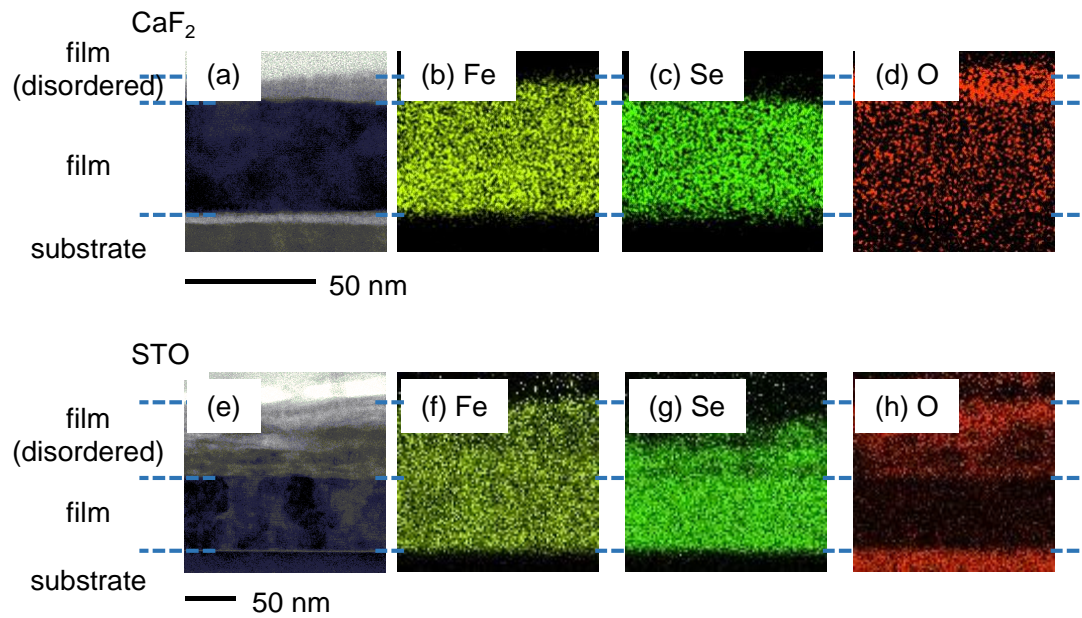

S. Figure 1 Topographic image of scanning TEM and EDX mapping images of Fe, Se and O. (a)-(d) correspond to FeSe$_{0.8}$Te$_{0.2}$ film on CaF$_2$ substrate. (e)-(h) correspond to FeSe$_{0.8}$Te$_{0.2}$ film on SrTiO$_3$ substrate.

## Thickness dependence of superconducting critical temperature for FeSe thin films on LAO and STO substrates

One FeSe film and three FeSe films were fabricated on LAO and STO substrates, respectively. The films were etched for $V_G$ of 5 V and the temperature dependence of the resistance was measured after the each etching cycle. Temperature dependence of the resistance is shown in S. Fig. 2(a) and (b). Critical temperature rapidly increased for the film on the LAO substrate. In contrast, the film on the STO substrate showed a $T_c$ increase with a thickness of around 10 nm. S. Figure 2(c) shows the thickness dependence of $T_c^{onset}$ for all FeSe samples.

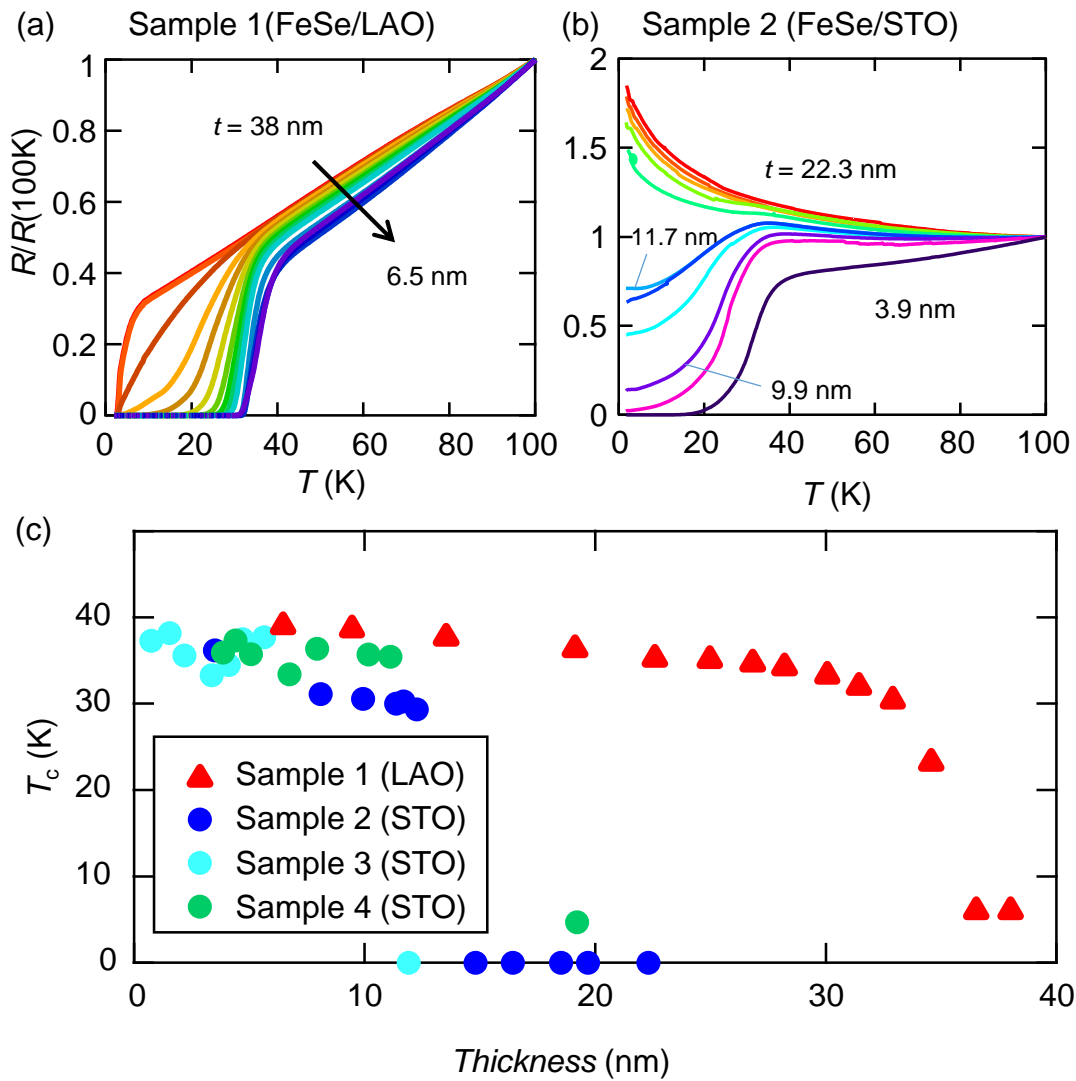

S. Figure 2 (a), (b) Temperature dependence of sheet resistance $R$ normalized by $R$ at 100 K with various thicknesses $t$. (c) Thickness dependence of superconducting critical temperature $T_c^{onset}$ for FeSe samples on LAO and STO substrates.

## Electrochemical reactions on the FeSe$_{0.8}$Te$_{0.2}$ / DEME-TFSI interface

For the thickness estimation, we assumed that the Faradaic charge $Q_F$ for each cycle is proportional to an etched thickness of the Fe(Se$_{0.8}$Te$_{0.2}$) film during the cycle. The observed $Q_F$ is sum of $Q_F$ by all electrochemical reactions in the ionic liquid. S. Table 1 shows $Q_F$ during an etching process of the FeSe$_{0.8}$Te$_{0.2}$ films and corresponding numbers of electrons per etched Fe atoms. 11.2 – 21.4 electrons flowed during etching of one Fe atom. The difference of the electron count probably originates from $Q_F$ by the side-reactions between the ionic liquid (or impurity in the ionic liquid) and electrodes. Since $Q_F$ by the side-reactions would be changed by amount of impurities in the ionic liquid and coverage of ionic liquid on the electrodes, $Q_F$ also showed the sample dependence.

Superconductivity has been reported on alkali-metal doped or (Li$_{0.8}$Fe$_{0.2}$)OH inserted FeSe with $T_c$ of 40 K. Then, we think that cation intercalated Fe(Se,Te) is a potential candidate of the conducting layer. For example, if DEME$^+$ ions in the ionic liquid are electrochemically inserted, following reaction would occur with $V_G$ of 5 V.(cathode reaction)

FeSe$_{0.8}$Te$_{0.2}$ + DEME$^+$ + e$^-$ -> FeSe$_{0.8}$Te$_{0.2}$(DEME)

In addition, since water is a major impurity in the ionic liquid, insertion of H$^+$ ions to (Se,Te) can also occur. When these reaction occurs, carrier density of the reacted layer increased by one electron/lattice.

However, an electrochemical etching of the FeSe$_{0.8}$Te$_{0.2}$ simultaneously occurs during the electrochemical doping. In addition, the observed electron count (>10 electrons) are much larger than electrons for the doping reaction, suggesting other electrochemical reaction. Since $V_G$ of 5 V is around the electrochemical window of the ionic liquid, DEME-TFSI, electrochemical decomposition of DEME$^+$ and/or TFSI$^-$ probably occurs, resulting in a formation of chemically reactive species. Indeed, the color of the ionic liquid after applying $V_G$=5V was changed to be yellowish. We think that such reactive species probably react with FeSe$_{0.8}$Te$_{0.2}$, resulting in the etching of the film.

In a water-based electrolyte, FeSe are electrochemically deposited by following raction.[S1]

FeSe + 4H$_2$O -> Fe$^{2+}$ + SeO$_4^{2-}$ + 8H$^+$ + 8e$^-$

Then, similar electrochemical decomposition of the surface conducting layer can occur by water in the ionic liquid for $V_G$ = 0 V.(anode reaction)

FeSe$_{0.8}$Te$_{0.2}$(DEME) + 4H$_2$O -> Fe$^{2+}$ + Se(Te)O$_4^{2-}$ +DEME$^+$ + 8H$^+$ + 9e$^-$.

Another possible reaction is dissolution of the surface conducting layer without oxidation of selenium,

FeSe$_{0.8}$Te$_{0.2}$(DEME) -> Fe$^{2+}$ + Se(Te)$^{2-}$ +DEME$^+$ + e$^-$.

Since the etching experiment was carried out at low temperature in vacuum, we cannot make detailed electrochemical measurements on this interface. For further understanding the results, investigations at room temperature with large FeSe0.8Te0.2 film electrode is needed.

| sample (substrate) | Faradaic charge $Q_F$ (mC) | Etched thickness $A$ (nm) | Channel area $S$ (mm$^2$) | Electrons per one Fe atom |
|---|---|---|---|---|
| sample 1(LAO) | 5.28 | 64.2 | 1.18 | 16.9 |
| sample 2 (CaF$_2$) | 5.76 | 74.3 | 1.00 | 18.9 |
| sample 3 (STO) | 6.16 | 63.9 | 1.09 | 21.4 |
| sample 4 (LAO) | 1.76 | 27.4 | 1.08 | 14.4 |
| sample 5(CaF$_2$) | 1.38 | 31.1 | 0.96 | 11.2 |

S. Table1. Faradaic charge during an etching process of the FeSe$_{0.8}$Te$_{0.2}$ films and corresponding numbers of electrons per etched Fe atoms. Samples 1-3 correspond to the devices shown in Fig. 3, and samples 4,5 correspond to the samples for the TEM measurements.

## Superconducting properties of the FeSe$_{0.8}$Te$_{0.2}$ films on LaAlO$_3$ substrate

We examined critical magnetic field for a pristine FeSe$_{0.8}$Te$_{0.2}$/LAO film and the surface conducting layer with $V_G$ =5 V (the sample is identical to the TEM sample shown in Fig. 4). As shown in S. Fig. 3, critical magnetic field of the surface conducting layer is higher than that of the pristine film.

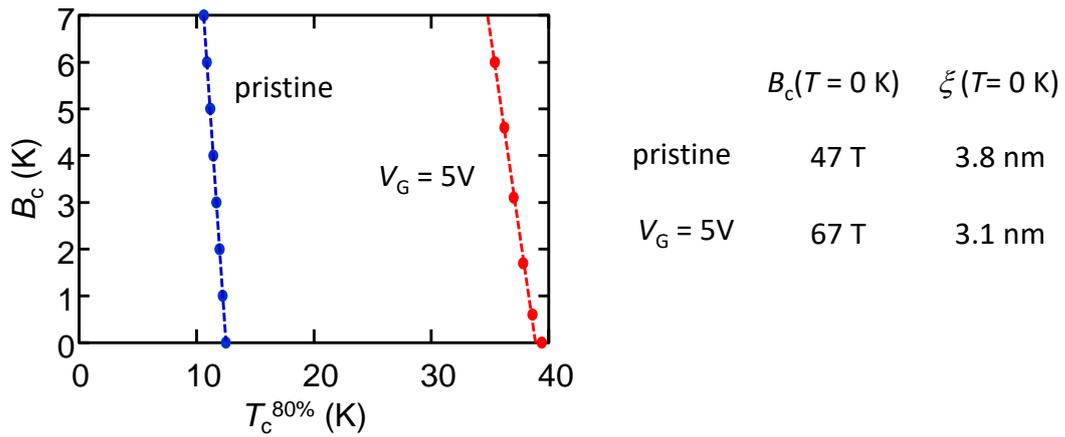

S. Figure 3 Temperature dependence of the critical magnetic field for the pristine film and the surface conducting layer with $V_G$ = 5V. The critical magnetic field and the coherence length at $T = 0$ K is shown in the right side.

When a thickness of a superconducting film is smaller than the superconducting coherence length $\xi$,

it behaves as a two dimensional superconductor. In that case, Berezinskii-Kosterlitz-Thouless (BKT) transition occurs below $T_c$. We analyzed the data in Fig. 2 (c) (FeSe$_{0.8}$Te$_{0.2}$/LAO with $V_G$ of 0 V and 5 V) with the Halperin-Nelson formula. [S2]

$$R_S = R_0 \exp\left(-2b\left(\frac{T}{T_0} - 1\right)^{\frac{1}{2}}\right)$$

This formula is changed to following equation,

$$\left(\frac{d\log(R_S)}{dT}\right)^{-2/3} = \left(\frac{b}{T_0}\right)^{-\frac{2}{3}}\left(\frac{T}{T_0} - 1\right)$$

When the superconductor is two dimensional, $T_0$ is much smaller than the critical temperature $T_c$, and $T_0$ equal to the BKT transition temperature $T_{BKT}$. As shown in S.Fig. 4(b), although $R$-$T$ curve seemed to be fitted by the formula, the transition temperature $T_0$ is almost identical to $T_c$. In addition, as shown in S. Fig. 4 (a), the fitted curve follows onset of the superconducting transition. These indicate that our sample did not show the BKT transition, and suggest that the thickness of the superconducting layer is larger than the coherence length.

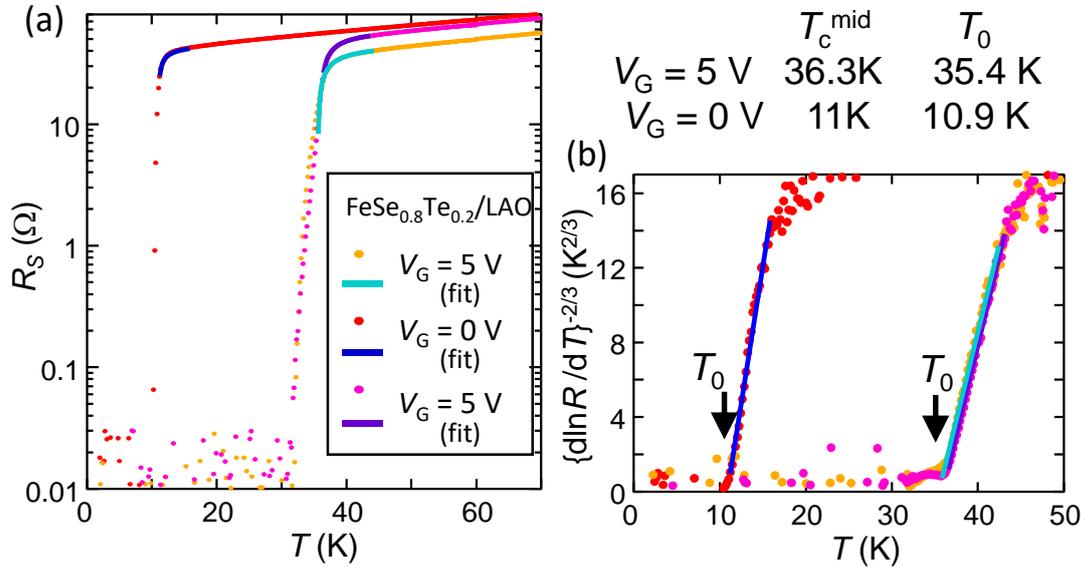

S. Figure4 Curve fitting by the Halperin-Nelson formula. (a) $R$-$T$ curves the fitted curve for the FeSe$_{0.8}$Te$_{0.2}$ film on LaAlO$_3$ substrate for $V_G$ of 0 V and 5 V. (c) Linear fitting of $\left(\frac{d\log(R_S)}{dT}\right)^{-2/3}$ as a function of temperature. $T_c^{mid}$ and $T_0$ are also shown.

## References

[S1] S. Demura, H. Okazaki, T. Ozaki, H. Hara, Y. Kawasaki, K. Deguchi, T. Watanabe, S. J. Denholme, Y. Mizuguchi, T. Yamaguchi, H. Takeya, Y. Takano, Solid State Commun. 154, 40 (2013).

[S2] Halperin, B. I. & Nelson, D. R. , J. Low Temp. Phys. 36, 599-616 (1979).